\documentclass[letter,longauth,traditabstract]{aa} 
\usepackage{graphicx}
\usepackage{txfonts}
\def\deg{\ifmmode{^\circ} \else {$^\circ$} \fi}
\def\arcmin{\ifmmode{^\prime} \else {$^{\prime}$} \fi}
\def\arcsec{\ifmmode{^{\prime \prime}} \else {$^{\prime \prime}$} \fi}

\begin{document}
   \title{
HERSCHEL-HIFI spectroscopy of the intermediate mass protostar NGC7129 FIRS 2
\thanks{HERSCHEL is an ESA space observatory with science instruments 
provided by European-led Principal Investigator consortia and with important 
participation from NASA.
}
}
\titlerunning{HIFI spectroscopy of NGC7129 FIRS 2}

\author{D. Johnstone\inst{1,2}
\and M. Fich\inst{3}
\and C. M$^{\textrm c}$Coey\inst{3,4}
\and T.A. van Kempen\inst{5,6}
\and A. Fuente\inst{7}
\and L.E. Kristensen\inst{5}
\and J. Cernicharo\inst{8}
\and P.~Caselli\inst{9,10}
\and R. Visser\inst{5}
\and R. Plume\inst{11}
\and G.J. Herczeg\inst{12}
\and E.F. van Dishoeck\inst{5,12}
\and S. Wampfler\inst{13}
\and R. Bachiller\inst{7}
\and A. Baudry\inst{13}
\and M. Benedettini\inst{14,15}
\and E. Bergin\inst{15}
\and A.O. Benz\inst{13}
\and P. Bjerkeli\inst{16}
\and G. Blake\inst{17}
\and S. Bontemps\inst{18}
\and J. Braine\inst{18}
\and S. Bruderer\inst{13}
\and C.~Codella\inst{14}
\and F. Daniel\inst{19,20}
\and A.M.~di~Giorgio\inst{14}
\and C. Dominik\inst{21}
\and S.D. Doty\inst{22}
\and P. Encrenaz\inst{23}
\and T. Giannini\inst{14}
\and J.R.~Goicoechea\inst{10}
\and Th.~de~Graauw\inst{24}
\and F. Helmich\inst{25}
\and F. Herpin\inst{18}
\and M.R. Hogerheijde\inst{5}
\and T. Jacq\inst{18}
\and J.K. J{\o}rgensen\inst{26}
\and B.~Larsson\inst{27}
\and D. Lis\inst{17}
\and R. Liseau\inst{16}
\and M. Marseille\inst{25}
\and G. Melnick\inst{6}
\and D. Neufeld\inst{30}
\and B. Nisini\inst{14}
\and M. Olberg\inst{16}
\and B. Parise\inst{28,29}
\and J. Pearson\inst{31}
\and C. Risacher\inst{25}
\and J. Santiago-Garc\'{i}a\inst{7}
\and P. Saraceno\inst{14}
\and R. Shipman\inst{25}
\and M. Tafalla\inst{7}
\and F. van der Tak\inst{25,32}
\and F.~Wyrowski\inst{28}
\and U.A. Y{\i}ld{\i}z\inst{5}
\and E. Caux\inst{32}
\and N. Honingh\inst{29}
\and W. Jellema\inst{25}
\and R. Schieder\inst{29}
\and D. Teyssier\inst{34}
\and N. Whyborn\inst{24}
}

\institute{
National Research Council Canada, Herzberg
Institute of Astrophysics, 5071 West Saanich Rd, Victoria, BC, V9E
2E7, Canada;
\email{doug.johnstone@nrc-cnrc.gc.ca}
\and
Department of Physics \& Astronomy, University of Victoria,
Victoria, BC, V8P 1A1, Canada
\and
Department of Physics and Astronomy, University of Waterloo,
Waterloo, Ontario, Canada N2L 3G
\and
Department of Physics and Astronomy, the University of Western Ontario, 
London, Ontario, Canada, N6A 3K7
\and
Leiden Observatory, Leiden University, P.O. Box 9513, 2300 RA Leiden, 
The Netherlands
\and
Harvard-Smithsonian Center for Astrophysics, 60 Garden Street, MS 42, Cambridge, MA 02138, USA
\and
Observatorio Astron\'{o}mico Nacional (IGN), Apartado 1143, 28800 Alcal\'{a} de Henares, Spain
\and
Department of Astrophysics, CAB, INTA-CSIC, Crta Torrej\'{o}n a Ajalvir km 4, 28850 Torrej\'{o}n de Ardoz, Spain
\and
School of Physics and Astronomy, University of Leeds, Leeds LS2 9JT, UK 
\and
INAF - Osservatorio Astrofisico di Arcetri, Largo E. Fermi 5, 50125 Firenze, Italy
\and
Department of Physics and Astronomy, University of Calgary,
Calgary, Alberta, Canada 
\and
Max-Planck-Institut f\"ur extraterrestrische Physik, Garching, Germany
\and
Institute of Astronomy, ETH Z\"urich, 8093 Z\"urich, Switzerland
\and
INAF - Istituto di Fisica dello Spazio Interplanetario, Area di Ricerca di Tor Vergata, via Fosso del Cavaliere 100, 00133 Roma, Italy
\and
Department of Astronomy, The University of Michigan, 500 Church Street, Ann Arbor, MI 48109-1042, USA
\and
Department of Radio and Space Science, Chalmers University of Technology, Onsala Space Observatory, 439 92 Onsala, Sweden
\and
California Institute of Technology, Division of Geological and Planetary Sciences, MS 150-21, Pasadena, CA 91125, USA
\and
Universit\'{e} de Bordeaux, Laboratoire d¿Astrophysique de Bordeaux, France; CNRS/INSU, UMR 5804, Floirac, France
\and
Observatoire de Paris-Meudon, LERMA UMR CNRS 8112, 5 place Jules Janssen, 92195 Meudon Cedex, France
\and
Department of Molecular and Infrared Astrophysics, Consejo Superior de Investigaciones Cientificas, C/ Serrano 121, 28006 Madrid, Spain
\and
Astronomical Institute Anton Pannekoek, University of Amsterdam, Kruislaan 403, 1098 SJ Amsterdam, The Netherlands 
\and
Department of Physics and Astronomy, Denison University, Granville, OH, 43023, USA
\and
LERMA and UMR 8112 du CNRS, Observatoire de Paris, 61 Av. de l'Observatoire, 75014 Paris, France
\and
Atacama Large Millimeter/Submillimeter Array, Joint ALMA Office, Santiago, Chile 
\and
SRON Netherlands Institute for Space Research, Landleven 12, 9747 AD Groningen, The Netherlands
\and
Centre for Star and Planet Formation, Natural History Museum of Denmark, University of Copenhagen,{\O}ster Voldgade 5-7, DK-1350 Copenhagen, Denmark
\and
Department of Astronomy, Stockholm University, AlbaNova, 106 91 Stockholm, Sweden
\and
Max-Planck-Institut f\"{u}r Radioastronomie, Auf dem H\"{u}gel 69, 53121 Bonn, Germany
\and
KOSMA, Physikalisches Institut, Universit\"{a}t zu K\"{o}ln, Z\"{u}lpicher Str. 77, 50937 K\"{o}ln, Germany
\and
Department of Physics and Astronomy, Johns Hopkins University, 3400 North Charles Street, Baltimore, MD 21218, USA
\and
Jet Propulsion Laboratory, California Institute of Technology, Pasadena, CA 91109, USA
\and
Kapteyn Astronomical Institute, University of Groningen, PO Box 800, 9700 AV, Groningen, The Netherlands
\and
Centre d'Etude Spatiale des Rayonnements, Universit\'e de Toulouse [UPS], 31062 Toulouse Cedex 9, France; 
CNRS/INSU, UMR 5187, 9 avenue du Colonel Roche, 31028 Toulouse Cedex 4, France
\and
European Space Astronomy Centre, ESA,ÊP.O. Box 78, E-28691 Villanueva de la Ca\~{n}ada, Madrid, Spain
 }

   \date{\bf }

  \abstract
   {
HERSCHEL-HIFI observations of water from the intermediate mass protostar NGC7129 FIRS 2 provide a powerful diagnostic of the physical conditions in this star formation environment. 
Six spectral settings, covering four H$_2$$^{16}$O and two H$_2$$^{18}$O lines, were observed and all but one H$_2$$^{18}$O line were detected. The four H$_2$$^{16}$O  lines discussed here share a similar morphology: a narrower, $\approx$ 6\,km\,s$^{-1}$, component centered slightly redward of the systemic velocity of NGC7129 FIRS 2 and a much broader, $\approx 25$\,km\,s$^{-1}$ component centered blueward and likely associated with powerful outflows. The narrower components are consistent with emission from water arising in the envelope around the intermediate mass protostar, and the abundance of H$_2$O is constrained to $\approx$10$^{-7}$ for the outer envelope.  Additionally, the presence of a narrow self-absorption component for the lowest energy lines is likely due to self-absorption from colder water in the outer envelope. The broader component, where the  H$_2$O/CO relative abundance is found to be $\approx$ 0.2, appears to be tracing the same energetic region that produces strong CO emission at high $J$.}
     \keywords{ Stars: formation - ISM: molecules}

   \maketitle
%

\section{Introduction}


Observations of star formation in nearby molecular clouds have provided a reasonably clear picture of this process for low-mass stars. Nevertheless fundamental questions remain. One of these is the variation in the formation process between low through high mass stars. The intermediate mass (IM) sub-program within the Water In Star-forming regions with HERSCHEL (WISH) key program (van Dishoeck et al \cite{vandishoeck}) aims to use observations of water, and complementary molecules, to probe the regions around several intermediate mass protostars - focusing on the structure of the envelope and the energetics of the outflow.  These sources are usually defined as young stellar objects (YSOs) with bolometric luminosities between 75 L$_\odot$ and $2\times 10^3$ L$_\odot$. A key goal for the longer term of this investigation is to place IM star formation in context with its low- and high-mass brethren.

An excellent example of an extremely young IM protostar is NGC7129 FIRS 2 (Eiroa et al 1998; Fuente et al \cite{fuente01},\ \cite{fuente05a}), located at a distance of $1260\pm50$\,pc from the Sun (Shevchenko \& Yakubov, \cite{shevchenko}).  With a luminosity of 500 L$_\odot$ and estimated stellar mass of 5 M$_\odot$, FIRS 2 lies near the middle of the IM luminosity range.  The protostar has produced a hot core (Fuente et al \cite{fuente05a}) but has no evidence for a large, well-developed disk - implying it is a young source. A powerful quadrupolar outflow, likely due to the superposition of two bipolar jets, is found very close to this source (Fuente et al \cite{fuente01}).

NGC7129 FIRS 2 has been the subject of significant investigation over the last year. A robust model for the enshrouding, and pre-natal, envelope has been proposed by Crimier et al (\cite{crimier}) utilizing all available far infrared ({\it Spitzer}) and submillimeter (JCMT, IRAM) brightness measurements. Additionally, Fich et al (2010) used HERSCHEL PACS observations of highly energetic CO (up to CO $J=$ 33-32 which arises from a level at an energy equivalent 3093\,K) to analyse the energetics associated with NGC7129 FIRS 2. They found that these CO lines are much brighter than expected from the envelope alone, revealing the need for additional heating beyond simple reprocessing of the protostellar radiation. A warm slab model with temperatures greater than 1000\,K, considered a proxy for shock heating in or along the surfaces of the outflow lobes, provided a much better fit to the CO observations. A lack of spectral resolution in the PACS data hindered further investigation into the connection between the CO lines and the outflow.

In this paper we present a first-look at the water spectra obtained toward NGC7129 FIRS 2 with the HIFI instrument (de Graauw et al \cite{degraauw}) on the HERSCHEL Space Observatory (Pilbratt et al \cite{pilbratt}). The wealth of spectrally resolved water lines observed provide powerful diagnostic measures for the envelope and the outflow. 


\section{Observations}

HERSCHEL HIFI was used to obtain data for this project during the HERSCHEL science demonstration phase and the HIFI priority science program, from 3--20 March, 2010. The observations were taken using the fast dual-beamswitch mode with a standard nod of 3\arcmin using receiver bands 1, 2, and 4. In total six spectral settings were selected, covering four H$_{2}$$^{16}$O transitions and two H$_{2}$$^{18}$O transitions, with integration times between 15 minutes and 1 hour. Two isotopologue lines of CO  were also located within these settings. CO 10--9 was also observed on June 15, 2010, with a 10 minute integration in band 5 using the fast dual-beamswitch mode.   The complete set of observed lines are shown in Table \ref{tabObs}. All the observations were taken toward NGC7129 FIRS2, at $21^h43^m1.7^s\ 66^\deg 03\arcmin 23\arcsec$ (J2000). Diffraction-limited beamsizes range from 20\arcsec -- 40\arcsec (equivalent to 25,000 -- 50,000 AU at the distance of NGC7129).

\begin{table}[h]
\caption{Observed H$_2$$^{16}$O, H$_2^{18}$O, and CO transitions.}
\label{tabObs}
\scriptsize
\begin{center}
\begin{tabular}{r c c r c c}
\hline \hline
Transition & $\nu$ & $E_{\rm u}/k_{\rm B}$ & Transition & $\nu$ & $E_{\rm u}/k_{\rm B}$ \\ 
           & (GHz) & (K) &  & (GHz) & (K) \\ \hline
p-H$_2$$^{16}$O 1$_{11}$--0$_{00}$         & 1113.34                    & \phantom{1}53.4  & o-H$_2$$^{18}$O 1$_{10}$--1$_{01}$ & \phantom{1}547.68 & \phantom{1}60.5 \\
p-H$_2$$^{16}$O 2$_{02}$--1$_{11}$        & \phantom{1}987.93 & 100.8                      &  p-H$_2$$^{18}$O 2$_{02}$--1$_{11}$             & \phantom{1}994.68 & 100.7 \\
p-H$_2$$^{16}$O 2$_{11}$--2$_{02}$        & \phantom{1}752.03 & 136.9                      &  C$^{18}$O\ \ \ \ \ \ \ 5--4 & \phantom{1}548.83 & 79.02 \\
o-H$_2$$^{16}$O 3$_{12}$--3$_{03}$        & 1097.37                     & 249.4                     & $^{13}$CO \  \ \ \ \ 10--9& 1101.35 &     290.79    \\
                                                                              &                                     &                                 & CO 10--9\tablefootmark{a} \ \ \ \ \      10--9 & 1151.99 & 304.2  \\
 \hline
\end{tabular}
\end{center}
\tablefoottext{a} Only a preliminary reduction of this observation has been performed, it is discussed only in Section 3.2
\end{table}

The data were pipelined using HIPE 3.0\footnote{HIPE is a joint development by the HERSCHEL Science Ground Segment Consortium, consisting of ESA, the NASA HERSCHEL Science Center, and the HIFI, PACS and SPIRE consortia.} (Ott 2010) and the data analysis was completed using both HIPE 3.0 and CLASS.  The results of the two analyses were found to be the same to better than twenty percent, where the largest uncertainties were found during the fitting of multiple components. Before the reduction, the chop positions were differenced to check for extended emission and none was detected. During the reduction, the H and V polarizations were compared to assess quality and then averaged. In most cases the two polarizations showed no significant variations. The  H$_2$$^{16}$O 3$_{12}$--3$_{03}$ line, however, showed both a lower peak intensity and line width for the V polarization, resulting in an 8\% difference in integrated intensity. For both polarizations, the overall morphology of the spectrum is similar, with the variation likely due to the extended nature of the source and slight offsets in the H and V beam positions. Finally, the WBS and HRS spectra were compared for consistency. Regions containing spurs were excluded from the analysis and a first order polynomial baseline was removed from each spectrum. The WBS data were ultimately smoothed to the goal resolution of 1.1\,MHz and the antenna temperature $T_A^*$ was converted to $T_{\rm mb}$ by applying a main beam efficiency of 0.74.

\begin{figure}
\centering
\includegraphics[width=9.2cm]{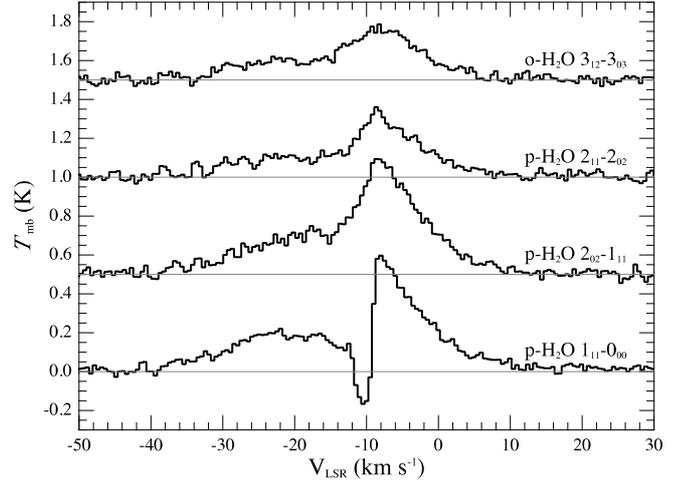}
\caption{
H$_2$O spectra detected toward NGC7129 FIRS 2 with HIFI. Note the similarities in morphology between the lines, especially the appearance of at least two emission components, narrow and broad (see Figure \ref{figfit}). The lowest energy transition lines also show evidence for somewhat blue-shifted self-absorption.}
\label{figh2o}
\end{figure}

\begin{figure}
\centering
\includegraphics[width=9.2cm]{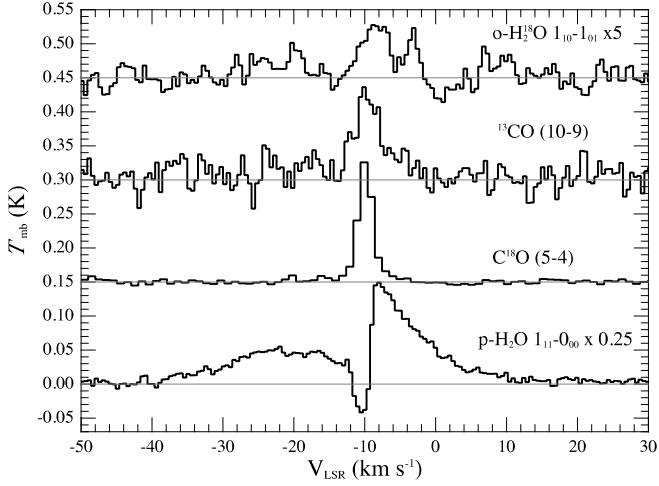}
\caption{
Isotopologue CO and water spectra observed toward NGC7129 FIRS 2 with HIFI. Note the width of the isotopologues relative to the main line.}
\label{figh2oco}
\end{figure}

The four H$_{2}$$^{16}$O transitions were observed and are shown in Figure \ref{figh2o}. The spectra show evidence for at least two components: a narrower line peaked slightly redward of the systemic velocity of the source ($-9.8$ km\,s$^{-1}$) and a broad line blue-shifted from the systemic velocity.  The ground-state transition shows significant self-absorption, with a minimum at $-10.4$\,km\,s$^{-1}$, hints of which can be seen in the other observed transitions, and is likely due to colder water lying along the line of sight toward NGC7129 FIRS 2, with the most likely source of absorption being the protostellar envelope.   Two CO isotopologues  were also detected (see Figure \ref{figh2oco}). While both these lines peak at the systemic velocity and are much narrower than the H$_2$$^{16}$O lines, the $^{13}$CO 10--9 line is broader than the C$^{18}$O 5--4 line  (see Table \ref{tablines}).   Of the water isotopologues, only the  H$_{2}$$^{18}$O 1$_{10}$  -- 1$_{01}$ line was sufficiently strong to be unambiguously detected (Figure \ref{figh2oco}).

\begin{table}[h]
\caption{Water and CO line emission toward NGC7129 FIRS 2}
\label{tablines}
\scriptsize
\begin{center}
\begin{tabular}{r r c c c c c}
\hline \hline
 Transition  & Component & rms\tablefootmark{a}  & $T_{\rm MB}^{\rm peak}$\tablefootmark{b} & $\int T_{\rm MB}~{\rm d}\varv$ \tablefootmark{b}& V$_{\rm LSR}$\tablefootmark{c} & FWHM\tablefootmark{d} \\
           & & (mK) & (K) & (K\,km\,s$^{-1}$) & (km\,s$^{-1}$) & (km\,s$^{-1}$)  \\
\hline
H$_2$$^{16}$O 1$_{11}$--0$_{00}$ & no Gaussian fit & 16 &  0.59& 8.4 &-8.2  &  \\
2$_{02}$--1$_{11}$ & narrow& 20 & 0.36 & 2.3 & -7.2  &   6.2\\
                                     & broad  & 20 & 0.23 & 6.1 & -12.4& 24.6\\ 
2$_{11}$--2$_{02}$ &narrow & 16 & 0.21 & 1.5 & -7.5 & 6.5\\
                                     &broad   & 16 & 0.11  & 3.1& -13.7& 26.8\\
3$_{12}$--3$_{03}$ &narrow & 16 & 0.14 & 1.7 & -7.3 & 7.8 \\
                                     &broad   & 16 & 0.11 & 2.8 & -14.5 & 24.4 \\
H$_2$$^{18}$O 1$_{10}$--1$_{01}$ &&  5 & 0.015  & 0.071 & -8.5 & 5.3 \\
H$_2$$^{18}$O 2$_{02}$--1$_{11}$ & not seen & 5  &  &$<$0.010\tablefootmark{e}  &  &  \\
C$^{18}$O\ \ \ \ \ \ \ 5--4  & &5    & 0.16 & 0.49 & -9.8& 2.1  \\
$^{13}$CO \  \ \ \ \ 10--9 & & 20 & 0.13 & 0.44 & -9.8& 5.5 \\
\hline
\end{tabular}

\end{center}
\tablefoottext{a}{The rms measure is taken over a 1 km\,s$^{-1}$ range.}\\
\tablefoottext{b}{Uncertainties in the peak temperature and integrated line intensity are dominated by calibration issues. For H$_2$$^{16}$O 1$_{11}$--0$_{00}$, 2$_{02}$--1$_{11}$, 3$_{12}$--3$_{03}$, and $^{13}$CO 10--9 the uncertainty is 32\%. For all other lines the uncertainty is 16\%. }\\
\tablefoottext{c}{Velocity accuracy better than $\sim$0.5 km\,s$^{-1}$.}\\
\tablefoottext{d}{A 20\% uncertainty in the line widths is due primarily to the component separation.}\\
\tablefoottext{e}{The 3$\sigma$ detection limit, for data smoothed to the expected line width (5\,km\,s$^{-1}$), is 0.015 K}\\
\end{table}

\begin{figure}
\centering
\includegraphics[width=9.2cm]{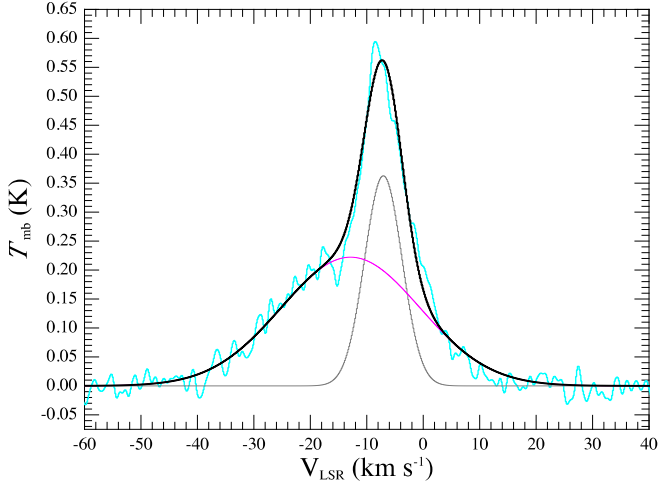}
\caption{
Spectrum of the H$_{2}$$^{16}$O $2_{02}$  -- 1$_{11}$ transition (blue) superposed with the two Gaussian fit (black) from Table \ref{tablines}. The fit includes a narrow, $6.2$\,km\,s$^{-1}$ FWHM, component centered at  -7.2\,km\,s$^{-1}$ (grey) and a much broader, $24.6$\,km\,s$^{-1}$ FWHM component, shifted slightly blue-ward at -12.4\,km\,s$^{-1}$ (purple).}
\label{figfit}
\end{figure}

\section{Discussion}

The H$_{2}$$^{16}$O lines detected toward NGC7129 FIRS 2 provide an opportunity to examine the physical and chemical environment of this young IM protostar.   Attempts to fit the spectra with two Gaussian components proved reasonably successful for the excited lines (Table \ref{tablines}  \& Figure \ref{figfit}). For these three lines, the narrower component was well described by a $\approx$ $6$ km\,s$^{-1}$, Gaussian centered redward of the systemic velocity of the source near $-7$ km\,s$^{-1}$ coupled with a broader, $25$ km\,s$^{-1}$, Gaussian centered near $-13.5$ km\,s$^{-1}$ (Figures \ref{figh2o} \& \ref{figfit}). The likely additional absorption component in each of these spectra could account for the small variations in the line centroids and widths of the fits.   Similar profiles are seen for low- (Kristensen et al. 2010) and high-mass (Chavarr\'ia et al. 2010) protostars, where they are named medium and broad components. Even narrower lines, \mbox{$\approx$ 1-2 \,km\,s$^{-1}$} FWHM, are seen in self-absorption of the lowest-energy water lines as well as in ground-based emission lines such as CS 3--2. The latter profile also reveals a 6\,km\,s$^{-1}$ component (Fuente et al. 2005).

Beyond the morphological nature of these lines, it is possible to compare the water emission against model predictions. Fich et al (\cite{fich}) note that NGC7129 FIRS 2 has two main components which may account for the observed emission. The prenatal envelope, which is warmed by the IM protostar embedded within it, may produce water emission. Also, the source powers an energetic outflow which could produce water emission either in the outflow itself or along the shock heated walls, where the outflow interacts with the denser envelope. 
In the following subsections, we consider the water emission from each of these components and compare the expectations against the HIFI observations.

\subsection{Water from a spherical protostellar envelope}

The spherical envelope model for NGC7129 FIRS 2 computed by Crimier et al (\cite{crimier}) fits the spectral energy distribution from the far infrared through the submillimeter very well and thus provides a reasonable set of physical conditions for examining the observed water emission.  This model has an envelope mass of 50 M$_\odot$, an optical depth at 100 $\mu$m of 2.3, an inner radius of \mbox{100 AU} and an outer radius of \mbox{18,600 AU}.  The temperature at the inner envelope radius is 289 K, falling to 100 K at a radius of 373 AU where the H$_2$ density is $4.4 \times 10^7$ cm$^{-3}$.  The density varies as a power-law with index -1.4. 

Model water lines were calculated using the RATRAN code (Hogerheijde \& van der Tak 2000) using collisional cross sections for H$_2$O-He from Green (1993), scaled by 1.348 to make a first order approximation for collisions with H$_2$. An  ortho-to-para ratio of 3 was assumed for H$_2$.  The density and temperature profile of the Crimier et al (\cite{crimier}) model were followed and the radial velocity profile was assumed to be the free-fall velocity appropriate to a central object of mass 1.1 M$_\odot$ (ie. velocity of -4.2 \,km\,s$^{-1}$ at the inner radius and -0.31 \,km\,s$^{-1}$ at the outer edge).   Models were calculated for two abundance zones in the envelope:  a warm, $>100$\,K inner envelope with a higher abundance and a cooler outer envelope with a lower abundance, perhaps characteristic of a ``freeze-out'' zone. 

As a starting point, we modelled the ortho-H$_{2}$$^{18}$O 1$_{10}$  -- 1$_{01}$ observation.  This line has a similar profile to the narrower components of the H$_2$O lines and if it can be reproduced by the spherical envelope model we may conclude that the narrower component of the H$_2$O lines also are likely to arise in the envelope.  A large number of models were run over a range of inner (10$^{-9}$--10$^{-4}$) and outer (10$^{-7}$--3$\times 10^{-11}$) envelope abundances.   The non-thermal velocity width parameter {\it b} was also varied and values \mbox{$\ge$ 3 \,km\,s$^{-1}$} were strongly ruled out,  while $b=1$\,km\,s$^{-1}$ produced too narrow a line.  The best fit was found for $b=2$\,km\,s$^{-1}$.  

The abundance of ortho-H$_{2}$$^{18}$O in the outer envelope is well-constrained to be $3\pm 1 \times 10^{-10}$, while inner envelope abundances between $3\times 10^{-7}$ and $1\times 10^{-5}$ produce results consistent with the integrated flux of the ortho-H$_{2}$$^{18}$O 1$_{10}$  -- 1$_{01}$ line (Table \ref{online1}).   These are surprisingly high values; indeed, the top end of this range is impossible when the cosmic $^{18}$O abundance is considered.   The para-H$_{2}$$^{18}$O 2$_{02}$  -- 1$_{11}$ was modeled in a similar way. The non-detection of this line constrains the outer envelope abundance of para-H$_{2}$$^{18}$O to $\le$ $3\times 10^{-11}$.  The model fits are insensitive to the inner envelope abundance used (Table \ref{online2}) and this may be a consequence of beam dilution.

Assuming that the ratio of $^{16}$O/$^{18}$O = 550 is also representative of the ratios of both of the ortho and para H$_{2}$O/H$_{2}$$^{18}$O, we examined models for the narrower components of the H$_2$O lines with outer envelope abundances near $1\times 10^{-7}$ and inner abundances up to $3\times 10^{-4}$ (limited by the cosmic O abundance).  The best fitting models indicate that the outer envelope abundance of total H$_2$O is of order a few 10$^{-7}$, and are summarised in Table \ref{tabfitsum}.  The ortho to para ratio implied by the best fit outer envelope abundances is 10:1, this is highly unlikely and within the calibration ($\le$32\%) and model uncertainties the ratio is consistent with 3.  The abundance of the inner envelope could not be constrained and the observed line profiles were poorly reproduced, particularly in the case of the para lines.  Whilst the parameter space of the models are still not fully explored, this suggests limitations in the model used with the assumptions concerning the velocity field and spherical structure most suspect.    

\begin{table}[h]
\caption{Overview of best fitting spherical envelope models}
\label{tabfitsum}
\scriptsize
\begin{center}
\begin{tabular}{r r l}
\hline \hline
Isotopologue & Outer abundance \tablefootmark{a}  & Comment \\
\hline
o-H$_2$$^{18}$O & $3 \pm 1 \times 10^{-10}$ & implies o-H$_2$O $1.7\,\pm\,0.5\times10^{-7}$ \tablefootmark{b}  \\
p-H$_2$$^{18}$O & $\le 3 \times 10^{-11}$      & implies p-H$_2$O $1.7\times10^{-8}$ \tablefootmark{b}   \\
o-H$_2$O               & $3 \times 10^{-7}$              & Well constrained by both excited para lines\\
p-H$_2$O               & $3 \times 10^{-8}$              & $\ge 10^{-8}$ required for self-absorption \\
\hline
\end{tabular}
\end{center}
\tablefoottext{a}{The relative abundance of species listed, and not total H$_2$O}\\
\tablefoottext{b}{Assuming $^{16}$O/$^{18}$O = 550  }\\
\end{table}

\subsection{Water from a slab model}

Fich et al \cite{fich} concluded that the energetic PACS CO lines observed toward NGC7129 FIRS 2 could not be fit by a reprocessing spherical envelope model as there was an insufficient volume of warm gas to produce the strong emission observed. It was postulated that the CO emission might arise along shock heated walls in the outflow cavity, where the gas temperature can reach extreme values. NGC7129 FIRS 2 is known to contain a powerful outflow (Fuente et al 2001) and observations of low-mass stars have shown that such outflows can produce heated walls via UV photons (van Kempen et al 2009) and shocks (Gianninni et al 1999, van den Ancker et al 2000, Nisini et al 2002, Arce et al 2007, van Kempen et al 2010).   Fich et al (\cite{fich}) used the RADEX code to show that the observed CO emission could be explained within a model of a slab geometry with a temperature of 1100\, K, an H$_2$ density of $1.0 \times 10^8\,$cm$^{-3}$, and a CO column density of $10^{14}$\,cm$^{-2}$. This model constrained the lowest temperature at which the CO observations could be adequately fit. A more comprehensive set of CO models was produced for this paper, and a best fit is found for a temperature of 1200\, K, an H$_2$ density of $10^7$ cm$^{-3}$ and a column density of CO of $1.7 \times 10^{14}$ cm$^{-2}$.  Assuming a CO abundance of $10^{-4}$, these results suggest a slab with thickness $dz \sim 10^{10}\,$cm.

Using these same conditions (1200\, K and $10^7$ cm$^{-3}$), but substituting water for CO, may provide an additional constraint on the possible shock heating conditions, and the conditions under which the broad component of the water lines arise.    The water lines observed with HIFI set the limit on the ortho-H$_2$O column at $2.3 \pm 0.2 \times 10^{13}$ cm$^{-2}$ while the para-H$_2$O lines observed set a maximum limit in the column at $7.5 \pm 0.5 \times 10^{12}$ cm$^{-2}$, giving an ortho-to-para ratio of 3.  

These models imply a water abundance that is approximately five times lower than CO and can be compared with that found from analysis of the H$_2$O and CO line ratios in the line wing, where the emission is optically thin.   The CO 10--9 line exhibits a similar profile to that seen in the H$_2$O lines (Figure \ref{online3}) and be compared directly with the \mbox{H$_{2}$O $2_{02}$  -- 1$_{11}$} line,  as they are observed in almost the same beam (22" vs 19").  The  abundance ratio was calculated with RADEX using a density of 10$^6$ cm$^{-3}$.  Above $T$= 150 K  the results are insensitive to changes in temperature.  The abundance ratio is found to be $\approx$ 0.2 near  the line centre and approaches 0.3 at the highest velocities in the red wing (Table \ref{online4}).  This is consistent with the findings from the slab model and also with the behaviour seen in the low-mass sources of Kristensen et al. (2010), who report this indicates that $\approx$ 10\% of the available O is in H$_2$O. 

That the same model can be used to fit both the PACS CO observations and the broad component of the HIFI H$_2$O lines supports the hypothesis that these lines arise along shock heated walls in the outflow cavity.  Shock models will allow to better constrain the conditions under which these broad lines arise.


\section{Conclusions}

HERSCHEL HIFI spectroscopy of water in the vicinity of the intermediate mass protostar NGC7129 FIRS 2 has revealed that the water emission arises from at least two sources. The observed emission lines can be decomposed into both a narrower,$\approx$  6\,km\,s$^{-1}$, component and a much broader, $\approx$ 25\,km\,s$^{-1}$ component. The integrated intensity of the narrower component can be fit by a simple free-falling envelope model with an outer envelope total H$_2$O abundance of $\approx$ 10$^{-7}$, although the shape
of the line profiles are not reproduced.  The broader component appears to be related to the heated gas already observed in high-$J$ CO lines with PACS and associated with the known energetic outflow.  These initial modeling results suggest that a dedicated parameter study for this source should prove extremely fruitful in constraining the physical and chemical conditions in NGC7129 FIRS 2.

\begin{acknowledgements}
We thank the HIFI ICC for all of their help with the data reduction, and both the referee and journal editor for critical comments and speed of response. 
JC and AF give thanks to Spanish MCINN for funding support under program
CONSOLIDER INGENIO 2010 ref: CSD2009-00038, and JC, under programs
AYA2006-14786 and  AYA2009-07304.
A portion of this research was performed at the Jet Propulsion Laboratory, California Institute of Technology, under contract with the National Aeronautics and Space Administration.
This program is made possible thanks to the HIFI guaranteed time program.
HIFI has been designed and built by a consortium of institutes and university departments from across Europe, Canada and the United States under the leadership of SRON Netherlands Institute for Space Research Groningen, The Netherlands and with major contributions from Germany, France, and the US. Consortium members are: Canada: CSA, U.Waterloo; France: CESR, LAB, LERMA, IRAM; Germany: KOSMA, MPIfR, MPS; Ireland: NUI Maynooth; Italy: ASI, IFSI-INAF, Osservatorio Astrofisico di Arcetri-INAF; Netherlands: SRON, TUD; Poland: CAMK, CBK; Spain: Observatorio Astronomico Nacional (IGN), Centro de Astrobiologia (CSIC-INT); Sweden: Chalmers University of Technology - MC2, RSS \& GARD, Onsala Space Observatory, Swedish National Space Board, Stockholm University - Stockholm Observatory; Switzerland: ETH Zurich, FHNW; USA: Caltech, JPL, NHSC.
\end{acknowledgements}

\Online

\begin{appendix}
\section{Online Material}

\begin{table}[h]
\caption{Summary of o-H$_2$$^{18}$O $1_{10}$ -- 1$_{01}$ model results.\tablefootmark{a}  }
\label{online1}
\scriptsize
\begin{center}
\begin{tabular}{cccc}
\hline \hline
\multicolumn{2}{c} {Envelope abundance} & $T_{\rm MB}^{\rm peak}$ & $\int T_{\rm MB}~{\rm d}\varv$ \\
Inner & Outer & (K) & (K km s$^{-1}$) \\
\hline
1$\times 10^{-5}$ &      3$\times 10^{-10}$ &   0.017  &  0.071  \\
3$\times 10^{-6}$ &      3$\times 10^{-10}$ &   0.018  &  0.075  \\
1$\times 10^{-6}$ &      3$\times 10^{-10}$ &   0.017  &  0.069  \\
3$\times 10^{-7}$ &      3$\times 10^{-10}$ &   0.016  &  0.065  \\
1$\times 10^{-7}$ &      3$\times 10^{-10}$ &   0.015  &  0.060  \\   
3$\times 10^{-8}$ &      3$\times 10^{-10}$ &   0.014  &  0.054  \\
1$\times 10^{-8}$ &      3$\times 10^{-10}$ &   0.014  &  0.050  \\
\hline
1$\times 10^{-5}$ &      2$\times 10^{-10}$ &   0.015  &  0.063  \\
3$\times 10^{-6}$ &      2$\times 10^{-10}$ &   0.014  &  0.060  \\
1$\times 10^{-6}$ &      2$\times 10^{-10}$ &   0.014  &  0.060  \\
3$\times 10^{-7}$ &      2$\times 10^{-10}$ &  0.013  &  0.052  \\
\hline
1$\times 10^{-5}$ &      1$\times 10^{-10}$ &   0.010  &  0.043  \\
1$\times 10^{-6}$ &      1$\times 10^{-10}$ &  0.009  &  0.038  \\
3$\times 10^{-7}$ &      1$\times 10^{-10}$ &   0.009  &  0.036  \\
1$\times 10^{-7}$ &      1$\times 10^{-10}$ &   0.008  &  0.032  \\
3$\times 10^{-8}$ &      1$\times 10^{-10}$ &   0.007  &  0.027  \\
\hline
\end{tabular}
\end{center}
\tablefoottext{a}{The observed line has a $T_{\rm MB}^{\rm peak}$ =  0.015 K and an $\int T_{\rm MB}~{\rm d}\varv$ =  0.071 K km$^{-1}$, the best fitting model is selected by comparison with the latter.}
\end{table}

\begin{table}[h]
\caption{Summary of p-H$_2$$^{18}$O $2_{02}$ -- 1$_{11}$ model results, which are compared with $\int T_{\rm MB}~{\rm d}\varv \le$  0.010 K km$^{-1}$}
\label{online2}
\scriptsize
\begin{center}
\begin{tabular}{cccc}
\hline \hline
\multicolumn{2}{c} {Envelope abundance} & $T_{\rm MB}^{\rm peak}$ & $\int T_{\rm MB}~{\rm d}\varv$ \\
Inner & Outer & (K) & (K km s$^{-1}$) \\
\hline
3$\times 10^{-6}$ &        1$\times 10^{-10}$ &      0.0088 &    0.033    \\
1$\times 10^{-6}$ &        1$\times 10^{-10}$ &      0.0088 &    0.033    \\
3$\times 10^{-7}$ &        1$\times 10^{-10 }$ &     0.0085 &    0.032    \\
\hline
3$\times 10^{-6}$ &         3$\times 10^{-11}$ &     0.0026 &   0.0097    \\
1$\times 10^{-6}$ &         3$\times 10^{-11}$ &     0.0027 &   0.0099    \\
3$\times 10^{-7}$ &         3$\times 10^{-11}$ &     0.0023 &   0.0085    \\
\hline
3$\times 10^{-6}$ &         1$\times 10^{-11}$ &     0.0008 &   0.0032    \\
1$\times 10^{-6}$ &         1$\times 10^{-11}$ &     0.0008 &   0.0031    \\
3$\times 10^{-7}$ &         1$\times 10^{-11 }$ &    0.0008 &   0.0030    \\
\hline
\end{tabular}
\end{center}
\end{table}

\begin{figure}
\centering
\includegraphics[width=9.2cm]{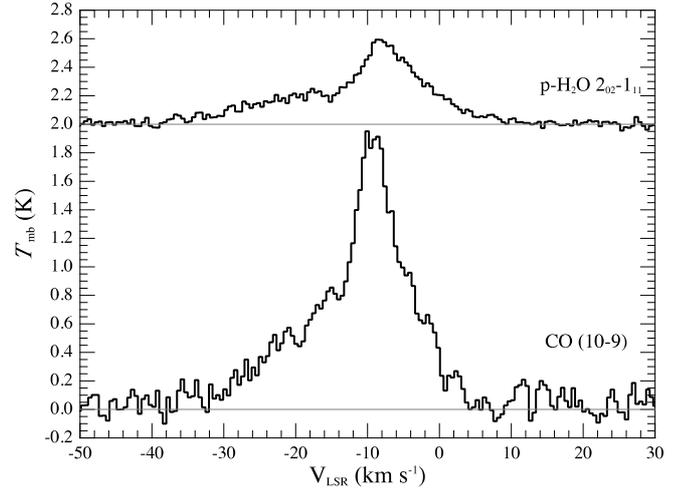}
\caption{
The H$_{2}$O $2_{02}$  -- 1$_{11}$ and CO 10--9 transitions; the main isotopologue of CO exhibits a profile similar to the main isotopologues of H$_2$O.}
\label{online3}
\end{figure}

\begin{table}[h]
\caption{H$_{2}$O $2_{02}$  -- 1$_{11}$/CO 10--9 abundance ratios in 5 km s$^{-1}$ intervals, calculated from  CO 10--9/ H$_{2}$O $2_{02}$  -- 1$_{11}$ line ratios, using  $n$ = 10$^6$ cm$^{-3}$ and $T > 150$ K}
\label{online4}
\scriptsize
\begin{center}
\begin{tabular}{rcc}
\hline \hline
$d\rm{v}_{\rm{LSR}}$ &  & \\
(km s$^{-1}$) &  CO 10--9/ & $x$(H$_2$O)/ \\
 &    H$_{2}$O 2$_{02}$  -- 1$_{11}$ & $x$(CO) \\     
 \hline
-10 -- -5 &  3.25 & 0.27 \\
-5 -- 0 &  3.86 & 0.22 \\
0 -- 5   &  4.14 & 0.21 \\
5 -- 10       & 1.86  &  0.47 \\
\hline
\end{tabular}
\end{center}
\end{table}

\end{appendix}

\end{document}